\documentclass[preprint,preprintnumbers,amsmath,amssymb]{revtex4}
\usepackage{graphicx}
\usepackage{dcolumn}
\usepackage{bm}

\begin{document}

\preprint{                }

\title{Reply to comment on ``Electronic structure and structural stability study of Li$_3$AlH$_6$'' }
\author{P. Vajeeston}
 \email{ponniahv@kjemi.uio.no}
\homepage{http://www.folk.uio.no/ponniahv}
\author{P. Ravindran}
\author{A. Kjekshus}
\author{H. Fjellv{\aa}g}
\affiliation{Department of Chemistry,
             University of Oslo, Box 1033, Blindern N-0315, Oslo, Norway}
\date{\today}
\begin{abstract}
The nature of the bonding in Li$_3$AlH$_6$ has been re-examined with
additional analyses using density-functional calculations. From
partial density of states, charge density distribution, charge
transfer, electron localization function, crystal orbital Hamilton
population and Mulliken population analyses it is concluded that the
interaction between Li and AlH$_6$ in Li$_3$AlH$_6$ is ionic as
earlier advocated. Based on charge density distribution, electron
localization function, and density of states analyses we earlier
suggested that the interaction between Al and H is largely of the
covalent type. However, additional analyses indicate that the
interaction between Al and H in the AlH$_6$ structural sub-units is of
a mixed covalent ionic character (iono-covalent).

\end{abstract}
\maketitle

The chemical bonding in materials becomes complicated to evaluate when
the number of constituents increases. For binary compounds one can
easily estimate the degree of ionic character from their
electronegativities using Gordy's~\cite{gordy55} approximation.  For
ternary and quaternary systems it is harder to estimate the bonding
character.  When we consider hydrides, it is even more complicated
than for other compounds, because of the small size of the hydrogen
and its only one valence electron which give rise to different bonding
character for the hydrogen in different chemical
environments (valence state +1 or $-$1 as well as covalent and
metallic character).  We have recently demonstrated these features in
a series of metal hydrides.\cite{raviprl,vajiprb} In our previous work
\cite{vajiprbr} on Li$_3$AlH$_6$, the main focus was on the structural
phase stability and electronic structure but we also characterized the
chemical bonding between the constituents based on results obtained
from partial density of states (PDOS), charge density and electron
localization function analyses. From these considerations we concluded
without ambiguity that the interaction between Li and the AlH$_6$
structural subunit is ionic. Energetic degeneration of Al-$p$ and
H-$s$ states (from DOS study), finite charge density distribution
between Al and H, and the polarization nature of ELF at the H site
toward Al, and the spatially constellation of Al and H led us to
conclude that the covalent bonding prevails between Al and H in the
AlH$_6$ subunit.  Moreover, the electronegativity difference between
Al and H is only 0.7, which also should favor covalent-type
interaction between them.

\par
A recent paper by Singh \cite{sing04} on Li$_3$AlH$_6$ describes the
bonding between H and Al as purely ionic.  The arguments for this
interpretation is based on DOS analysis and long-range Coulomb
interactions according to calculations using the linearized augmented
plane wave method. As more H-$s$ states are present in the valence
band (VB) than in the conduction band (CB), Singh concluded that the
interaction between Al and H must be ionic.  Aguayo and Singh
\cite{aguayo04} have also performed a similar type of analysis on
NaAlH$_4$ where a similar conclusion was reached (viz ionic
interaction between H and Al) The calculated DOS for Li$_3$AlH$_6$ by
Singh is in perfect agreement with our findings.\cite{vajiprbr} Hence,
there is no ambiguity between the different computational methods,
whereas the interpretation/understanding of the results differs.  It
is commonly recognized that it is difficult to characterize the nature
of chemical bonding (in particular for hydrides) from DOS and
integrated charges inside spheres alone. We have recently shown that
on using a combination of charge density, charge transfer and ELF
distribution along with other information one should be able to
characterize the quite complicated chemical bonding in
hydrides.\cite{raviprl} Hence in order to make a firm conclusion about
nature of the chemical bond in Li$_3$AlH$_6$ we need more data from
different perspectives. Hence, we have made additional calculations
using advanced density-functional tools.

\par
Each and every theoretical tool has some additional
flexibility/facility to evaluate bonding behavior. Hence we have now
used different density-functional tools to gather as much information
as possible regarding the chemical bonding in Li$_3$AlH$_6$.  The
calculations are made for the theoretical equilibrium structure
parameters specified in Ref. \onlinecite{vajiprbr} for Li$_3$AlH$_6$,
and for the compounds chosen for test cases we have used experimental
structural parameters. The DOS, charge density, charge transfer,
electron localization functions are evaluated from Vienna {\it ab
initio} simulation package,\cite{vasp} the crystal orbital population
is evaluated using the TBLMTO-47 package.\cite{tblmto} The Mulliken
population analyses have been made with the help of CRYSTAL03
\cite{crystal} code in which we used 5-11G, 6-11G, 5-11G, 8-61G, and
85-11G basis sets for H, Li, Be, Mg, and Al, respectively.

\par
\begin{figure}
\includegraphics[scale=0.425]{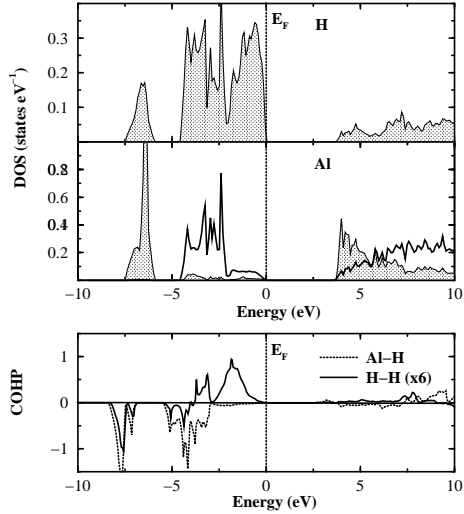}
\caption{Upper panels: Calculated site- and orbital-projected density of states for Al and H in Li$_3$AlH$_6$; $s$ electron contributions are shaded by grey color. Lower panel: COHPfor Al$-$H and H$-$H in Li$_3$AlH$_6$.}
\label{fig:pdos}
\end{figure}

The calculated partial DOS of Al and H in Li$_3$AlH$_6$ are shown in
Fig. \ref{fig:pdos}, illustrating the following three main features:
(1) The VB and CB are separated by a band gap of $\sim$3.8 eV
confirming that this compound is an insulator. (2) The VB is split
into two separate region by a $\sim$ 1.3 eV energy gap. The lowest
lying bands (at $-$7.5 to $-$5.8 eV) are mainly originated from Al $s$
with finite contributions from H-$s$ states.  The second region from
$\sim$ $-$4.5 to $-$2.2 eV comprises energetically degenerate H-$s$
and Al $p$ states which we focused on as a favorable situation for
formation of a covalent bonding in Ref. \onlinecite{vajiprbr}.  The
spherically symmetric nature of $s$ orbitals together with the
energetic degeneration of H-$s$ and Al-$s$ states, there should be a
high probability for formation of covalent-type bonding between these
atoms.  (3) The very small contribution of H in the unoccupied states
above the Fermi level (E$_F$) is explained with the help of COHP
analysis.  The COHP, which is the Hamiltonian population weighted DOS,
is identical to the crystal orbital overlap population. Negative value
of COHP indicates bonding character and the positive value of COHP
shows anti-bonding character. The DOS and COHP are evaluated using
different computer codes and that is the reason for the slight energy
shift between the DOS and COHP curves in Fig.\ref{fig:pdos}, but the
overall features are the same. The lower panel in Fig. \ref{fig:pdos}
(lower panel) shows that bonding states are present below $-$6.5 eV
and anti-bonding states between $\sim$ $-$2.8 eV and E$_F$. This
explains why there are few H-$s$ states present in the CB.

\par
\begin{figure*}
\includegraphics[scale=0.65]{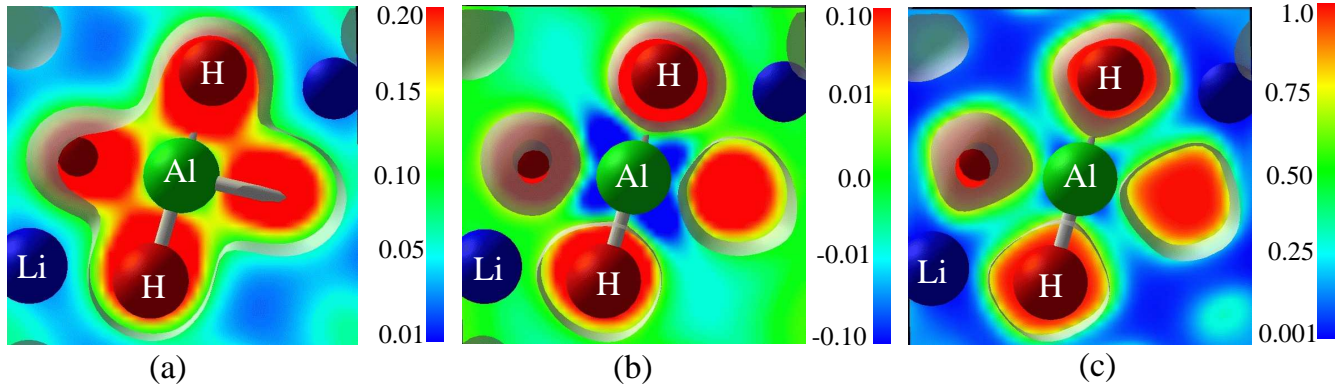}
\caption{(Color online) Calculated (a) valence-electron charge density,
(b) charge transfer, and (c) ELF plot for $\alpha$-Li$_3$AlH$_6$.} 
\label{fig:charge}
\end{figure*}

If the chemical bonding between Al and H is purely ionic one would
expect that Al-$p$ and H-$s$ states should be energetically well
separated.\cite{gellatt83} Further, in a purely ionic situation one
does not expect a finite electron density distribution between Al and
H as Fig. \ref{fig:charge}a shows.  Moreover the charge-density
distribution should be spherically symmetric around the H site if the
bonding is purely ionic.  Hence we came to our initial conclusion that
the bonding between Al and H in Li$_3$AlH$_6$ is largely covalent.
For strong covalent bonding Singh \cite{sing04} points out that the H
character should be distributed between the three $s$-$p$
manifolds. As the H-$s$ character is very small in the CB
(Fig. \ref{fig:pdos} upper panel), Singh concluded that covalent
contribution is negligible. However, in the first place one can not
judge the character of chemical bonding in complex materials based on
DOS analyses alone (e.g. see Ref.~\onlinecite{raviprl}).  Owing to the
iono-covalent interactions all H-$s$ orbitals will be filled and hence
both bonding and antibonding states of the $s$-$p$ hybrid are within
the VB as evident from our COHP analysis (e.g., see
Fig.\,\ref{fig:pdos} lower panel).

If the bonding interaction between all constituents in Li$_3$AlH$_6$
is purely ionic one would expect narrow band features and certainly
not the broad DOS features found for Li$_3$AlH$_6$ which indicate
overlap interaction between the constituents, viz. also for the purely
ionic case the distinct DOS manifold around $-$7 eV with 2 electrons
per formula unit are contributed by only one of the
constituents. However, considerable amount of electrons (according to
the integrated DOS which shows that this contribution is 36\% from Al,
49\% from H and the remaining 15\% from Li) from both Al-$s$ and H-$s$
indicate a finite degree of covalent character.  If Al is in the 3+
state, negligible amounts of electrons would be left at the Al site
resulting in very small contribution from it to the VB.  In fact, the
integrated DOS yielded $\sim$0.78 electrons at the Al site and shows
that the bonding interaction is not purely ionic.  The insulating
behavior can be explained as follows: Within one Li$_3$AlH$_6$ unit,
the 3 electrons from Li will fill 3 of the 6 half-filled H-$s$
orbitals and the remaining 3 half-filled H-$s$ orbitals form covalent
interaction with the three electrons from Al resulting completely
filled VB which gives the material insulating behavior.

In order to understand the chemical bonding in Li$_3$AlH$_6$ in detail
Singh \cite{sing04} performed a test calculation with the position of
one Li in the unit cell exchanged with a H.  Such an analysis is
questionable because although Li and H have one $s$ electron in the
outermost shell the former always takes the 1+ valence state whereas
the latter in the ionic case takes either 1+ or 1$-$ depending on the
chemical environment.

\par 
In order to gain further understanding of the nature of the bonding we
have calculated the charge density, charge transfer, and electron
localization function (ELF) for $\alpha$-Li$_3$AlH$_6$. The results
are shown in Fig. \ref{fig:charge}, where we show only the relation
between Al and H since there is no ambiguity with regard to the ionic
interaction between Li$^+$ and AlH$_6$$^-$.  The charge-density
distribution and charge-transfer plot indicate that a finite number of
electrons are present between Al and H, which means that there is a
finite covalent type of interaction between Al and H within the
AlH$_6$ unit. The charge transfer plot clearly indicates that charges
are depleted from Al (Fig. \ref{fig:charge}b) and Li. The depletion is
not spherically symmetric at the Al site, which implies that the
bonding is ionic with an appreciable covalent-type interaction
(directional bonding) between Al and H.  ELF is another useful tool to
distinguish between different bonding situations in
solids\cite{becke90,savin91}. The value of ELF is limited to the range
0 to 1. A high value of ELF corresponds to a low Pauli kinetic energy,
as can be found for covalent bonds or lone electron pairs.  The
calculated ELF for Li$_3$AlH$_6$ is shown in Fig. \ref{fig:charge}c in
which the H electrons are polarized toward the Al site. A similar
character of the ELF is found for the molecules C$_2$H$_6$ and
C$_2$H$_4$, where the interaction between the C and H is commonly
recognized as covalent.\cite{savin91} Hence, one must conclude that
there is a directional bonding character between Al and H in
Li$_3$AlH$_6$.

\par
In order to make a quantitative conclusion it would be useful to be
able to identify the amount of electrons on a particular atom and
populations between atoms. Although there is no unique definition of
how many electrons that are associated with an atom in a molecule or a
sub-unit of a solid it has nevertheless proven useful in many cases to
perform population analysis. Due to its simplicity the Mulliken
\cite{mulliken} population scheme has become the most familiar
approach to count the electrons associated with a given atom.  However
also this method is more qualitative than quantitative, giving results
that are sensitive to the atomic basis. Mulliken charges are reported
in Table 1 for series of H-based test materials using examples which
may provide benchmarks for systems with well recognized chemical
bonding.  LiH is a purely ionic compound and the calculated Mulliken
charges reflect nearly pure ionic picture with Li$^+$ and H$^-$. Also,
the overlap population between Li$^+$ and H$^-$ is close to zero, as
expected for an ionic compound.  Similarly in MgH$_2$, BeH$_2$, and
AlH$_3$ the bonding interaction is mainly ionic but the degree of
ionicity is reduced from MgH$_2$ to BeH$_2$ and further to AlH$_3$ viz
these compoundes exhibit some covalent character as evidenced by the
non zero overlap population. For the CH$_4$ molecule the
overlap population takes a value of 0.384 consistent with the well
known covalent interaction for this molecule. The Mulliken effective
charges for Li, Al and H in LiAlH$_4$ and Li$_3$AlH$_6$ indicate that
the interaction between the Li and AlH$_4$/AlH$_6$ is ionic (one
electron transfered from Li to AlH$_4$/AlH$_6$). There is a finite
overlap population between Al and H within the AlH$_4$/AlH$_6$ units
which reflects a partly covalent character of the Al$-$H
bond. However, the magnitude of overlap population is smaller than for
purely covalent compounds. Also, the partial charges (around two
electrons transfered from Al to H) implies that significant ionic
contribution to the Al$-$H bond.  The calculated ICOHP indicates that
the covalent Al$-$H interaction in LiAlH$_4$ is stronger than that in
Li$_3$AlH$_6$. Similarly, the calculated Mulliken effective charges
and overlap population indicates that the covalent Al$-$H interaction
is reduced when we move from LiAlH$_4$ to Li$_3$AlH$_6$.

\par
As a conclusion we thus find that the bonding nature of the hydrides
LiAlH$_4$ and Li$_3$AlH$_6$ do not exhibit simple ionic or covalent
character. In fact the bonding interaction in these compounds is quite
complicated.  The interaction between Li and AlH$_4$/AlH$_6$ is ionic
and that between Al and H comprises of ionic and covalent
character. We believe that similar type of bonding situation prevails
in all similar hydrides, but the magnitude of the ionic/covalent
mixture will exhibit considerable individual variation. Our study
indicates that several analyses must be performed in order to make a
finite conclusion regarding the bonding nature of such materials.

\begin{table}
\caption{Mulliken population analysis for selected hydrogen containing compounds. 
The Mulliken effective charges (MEC) are given in terms of $e$.} 
{
\scriptsize
\begin{ruledtabular}
\begin{tabular}{l l l c }
Compound & Atom & MEC & Overlap population   \\ \hline
LiH & Li & $+$0.98 & $-$0.003 (Li$-$H) \\
    & H  & $-$0.98 & \\
CH$_4$ & C & $-$0.26 & 0.384 (C$-$H) \\
    & H  & $+$0.06 & \\
MgH$_2$ & Mg & $+$1.87  & $-$0.040 (Mg$-$H) \\
    & H  & $-$0.93 &  \\
BeH$_2$ & Be &  +1.63 & 0.045 (Be$-$H) \\
    & H  & $-$0.82 & \\
AlH$_3$ & Al & +2.22  & 0.091 (Al$-$H) \\
    & H  & $-$0.74  & \\
LiAlH$_4$ & Li & +1.01 & 0.171 (Al$-$H) \\
    & Al & +2.01 & $-$0.021 (Li$-$H) \\
    & H  & -0.75 &   \\
Li$_3$AlH$_6$  & Li & +1.01 & 0.105 (Al$-$H) \\
         & Al & +2.08  & $-$0.020 (Li$-$H) \\
         & H  & $-$0.85   &  \\ 
\end{tabular}
\end{ruledtabular}
}
\label{table:str}
\end{table}

\par  
The authors gratefully acknowledge the Research Council of Norway for
financial support and for the computer time at the Norwegian supercomputer
facilities.

\end{document}